\documentclass[aps,prb,twocolumn,showpacs,showkey,square,numbers,amssymb,amsmath,nobibnotes]{revtex4-1}
\usepackage{bm,float}
\usepackage{times}
\usepackage{graphicx}
\usepackage[usenames,dvipsnames,svgnames,table]{xcolor}
\usepackage{hyperref}
\hypersetup{colorlinks=true, linkcolor=NavyBlue, citecolor=PineGreen,urlcolor=cyan}

\newcommand{\bracket}[1]{\left\langle #1\right\rangle}
\newcommand{\beeq}[1] {\begin{equation}\begin{split}#1\end{split}\end{equation}}

\begin{document}
\title{Level compressibility for the Anderson model on regular random graphs {\bf and the eigenvalue statistics in the extended phase}}
\author{Fernando L. Metz}
\address{Departamento de F\'isica, Universidade Federal de Santa Maria, 97105-900 Santa Maria, Brazil}
\author{Isaac P\'erez Castillo}
\address{Department of Complex Systems, Institute of Physics, UNAM, P.O. Box 20-364, 01000 Mexico City, Mexico}
\address{Department of Quantum Physics and Photonics, Institute of Physics, UNAM, P.O. Box 20-364, 01000 Mexico City, Mexico}
\address{London Mathematical Laboratory, 14 Buckingham Street, London WC2N 6DF, United Kingdom}

\begin{abstract}
We calculate the level compressibility $\chi(W,L)$ of the energy levels inside $[-L/2,L/2]$ for the Anderson model on infinitely large random regular graphs   with on-site potentials distributed uniformly in $[-W/2,W/2]$.   We show that $\chi(W,L)$ approaches the limit $\lim_{L \rightarrow 0^+} \chi(W,L) = 0$ for a broad interval of the disorder strength $W$ within the extended phase, including the region of $W$ close to the critical point for the Anderson transition. These results strongly suggest that the energy levels follow the Wigner-Dyson statistics in the extended phase, consistent with earlier 
analytical predictions for the Anderson model on an Erd\"os-R\'enyi random graph. Our results are obtained from the accurate numerical solution of an exact set of equations  valid for infinitely large regular random graphs.
\end{abstract}
\maketitle

\section{Introduction}

Despite more than fifty years since the seminal work of Anderson \cite{Ander58}, the localization of single particle wavefunctions in disordered  quantum systems continues to attract a significant interest \cite{abr2010}. Exactly solvable models naturally have played a crucial role in the understanding of the transition between localized and extended wavefunctions. One of the most important models consists of a single particle hopping among the nodes of an infinitely large Cayley tree with on-site disorder \cite{Abou73}.  In contrast to its counterpart in finite dimensions, this mean-field version of the Anderson model is exactly solvable thanks to  the absence of loops.

The statistics of energy levels and eigenfunctions of the Anderson model on locally treelike random graphs have lately re-emerged as a central problem in condensed matter theory. This is due to the connection between this class of models and localization in interacting many-body systems.  Essentially, the structure of localized wavefunctions in the Fock space of many-body quantum systems can be mapped on the localization problem of a single particle hopping on a tree-like graph with quenched disorder \cite{Altshuler1997,Gornyi2005,Basko2006}.  The phenomena of many-body localization and ergodicity breaking in isolated quantum systems prevent them to equilibrate, which has serious consequences for the foundations of equilibrium statistical mechanics \cite{Huse2015,Pino2016}.

The prototypical model to inspect the statistics of energy levels  and  eigenfunctions in the Anderson model  is realized on a regular random graph (RRG). Regular random graphs have a local treelike structure, but loops containing typically $O(\ln N)$ sites are present. Another difference of a RRG with respect to a Cayley tree (both with finite $N$) is the absence of boundary nodes in the former case. The majority of sites of a Cayley tree lies on its boundary, which pathologically influences the eigenfunctions within the delocalized phase \cite{Monthus,Tikhonov2016}.  Although the complete characterization of the phase diagram of the Anderson model on a RRG is still a work in progress \cite{Biroli2010,Aizenman1,Aizenman2}, it is well established that the extended phase appears at the center of the band as long as the disorder strength $W$ is smaller than a critical value $W_c$ \cite{Abou73,Biroli2012}.

Recently there has been an intense debate concerning the ergodicity of the eigenfunctions  within the extended phase of the Anderson model on RRGs and two main pictures have emerged. At one side, it has been put forward that, for a certain interval $W_E < W < W_c$, there exists an intermediate phase where the eigenfunctions are multifractal \cite{Biroli2012,Luca2014,Alt2016,Altshuler2016} and the inverse participation ratio scales as $N^{-\tau(W)}$ ($N \gg 1$), with $0 < \tau(W) < 1$ \cite{Alt2016,Altshuler2016}.  The results supporting this picture are mostly based on a numerical diagonalization study of the fractal exponents \cite{Luca2014,Alt2016,Altshuler2016}, combined with a semi-analytical  approach to solve the self-consistent equations \cite{Abou73}. The transition between ergodic and non-ergodic extended eigenstates at $W_E$  is discontinuous \cite{Alt2016,Altshuler2016}, analogous somehow to the one-step replica symmetry breaking transition observed in some spin-glass systems \cite{MezPar,MezPar1}. 

According to the other side, the inverse participation ratio scales as $1/N$ ($N \gg 1$) and the energy-levels follow the Wigner-Dyson statistics within the whole extended phase \cite{Mirlin2016,Garcia2016,Tikhonov2016}. The results supporting the ergodicity of the extended eigenstates are mainly based on numerical diagonalization \cite{Mirlin2016,Tikhonov2016}. The statistical properties of the energy levels and eigenfunctions display a non-monotonic behavior for increasing $N$, reducing essentially the non-ergodicity of the eigenfunctions to  a finite size effect. This picture is consistent with earlier analytical predictions for the problem of a single quantum particle hopping on an Erd\"os-Renyi random graph  \cite{Mirlin91,Fyod}, a treelike model closely related to the Anderson model on a RRG.

In this work we add an important contribution to this heated debate. We probe the eigenvalue statistics by solving an exact set of equations for the level compressibility $\chi$ of the number of energy levels inside the interval $[-L/2,L/2]$. By considering the limit $L \rightarrow 0^{+}$ (see below), this quantity allows to distinguish among the three conventional statistical behaviors of the energy levels found in Anderson models: Wigner-Dyson statistics \cite{BookMehta}, Poisson statistics (localized states) \cite{BookMehta} and sub-Poisson statistics (multifractal states)  \cite{Altshuler88,Chalker1996,Bogomolny2011,Mirlin}.  We calculate $\chi$ as a function of $L \ll 1$ within the extended phase, including some values of $W < W_c$ in an interval of disorder strength close to the critical point. This is the relevant regime of $W$ where one would expect the presence of multifractal eigenstates, according to recent numerical results \cite{Alt2016}. Our results consistently show that $\chi$ approaches zero in the limit $L \rightarrow 0^{+}$ for all values of $W < W_c$ considered here, which strongly supports the Wigner-Dyson statistics of the energy levels in the whole extended phase. 

The level-compressibility has a non-monotonic behavior as a function of $L$, which resembles the system size dependency discussed in previous works \cite{Mirlin2016,Alt2016}. Our approach is based on the numerical solution of an exact set of equations derived previously through the replica-symmetric method and valid in the limit $N \rightarrow \infty$ \cite{Metz2016}. We discuss the possible role of replica symmetry breaking in $\chi$ and the relation of our results with the problem of the existence of non-ergodic extended states.

The paper is organized as follows. In the next section we define the Anderson model on a regular random graph. We present the main equations for the level
compressibility and the corresponding results within the extended phase in section \ref{results}. Section \ref{discussion} discusses the possible role of replica
symmetry breaking and the relation of our results with previous works. The details of the numerical procedure to solve the equations for $\chi$ are presented in the appendix. 

\section{The Anderson model on a regular random graph.}  

The tight-binding Hamiltonian for a spinless particle moving on a random potential is given by
\beeq{
\mathcal{H}=-\sum_{ij = 1}^N t_{ij}\left(c_{i}^\dag c_{j}+c_{j}^\dag c_i \right)+\sum_{i=1}^N\epsilon_i c_i^\dag c_i\,,
\label{eq1aa}
}
where $t_{ij}$ is the energy for the hopping between nodes $i$ and $j$, while $\epsilon_1, \dots,\epsilon_N $ are the on-site random potentials drawn from the uniform distribution $P_{\epsilon}(\epsilon)=(1/W)\Theta(W/2-|\epsilon|)$, with $W \geq 0$. The hopping coefficients  $\{ t_{ij} \}_{i,j=1,\dots,N}$ correspond to the entries of the adjacency matrix of a regular random graph (RRG) with connectivity $k+1$ \cite{Bollobas,Wormald}. A matrix element $t_{ij}$ is equal to one if there is a link between nodes $i$ and $j$, and $t_{ij} = 0$ otherwise. The ensemble of random graphs can be defined through the full distribution of the adjacency matrix elements $\{ t_{ij} \}_{i,j=1,\dots,N}$. For the explicit form of this distribution, we refer the reader to \cite{Metz2016}.  For $k=2$, where each node is connected precisely to three neighbors, all eigenfunctions become localized provided $W > W_c \simeq 17.4$ \cite{Abou73,Biroli2010}. The value of $W_c$ is the same for the infinitely large Cayley tree and the RRG. \\

\section{Results for the level compressibility} \label{results}

Let $\mathcal{I}_L^{(N)}$ denotes the number of energy levels inside $[-L/2,L/2]$ 
\begin{equation}
 \begin{split}
\mathcal{I}^{(N)}_L= N\int_{-L/2}^{L/2} dE \, \rho_N(E)\,,
\label{ssq1}
\end{split}
\end{equation}
where $\rho_N(E)=(1/N)\sum_{i=1}^N\delta(E-E_i)$ is the density of energy levels $E_1,\ldots,E_N$ between $E$ and $E + dE$. We define the $N \rightarrow \infty$ limit of the level-compressibility as follows \cite{BookMehta,Mirlin}
\begin{equation}
\chi(L,W) = \lim_{N \rightarrow \infty} \frac{n^{(N)}(L) }{\left\langle \mathcal{I}^{(N)}_L  \right\rangle } \,,
\label{hhj}
\end{equation}
with the number variance
\begin{equation}
n^{(N)}(L) = \left\langle \left( \mathcal{I}^{(N)}_L \right)^2  \right\rangle  - \left\langle \mathcal{I}^{(N)}_L  \right\rangle^2
\end{equation}
characterizing the fluctuation of the energy levels within $[-L/2,L/2]$. The symbol $\langle \dots \rangle$  represents the ensemble average with respect to the graph distribution and the distribution of the  on-site potentials. 

Let us consider the behavior of $\chi(L,W)$ when $L=s/N$, i.e., the interval $[-L/2,L/2]$ is measured in units of the mean level spacing $1/N$. Energy levels following the Wigner-Dyson statistics strongly repel each other and the number variance scales as $n^{(N)}(L) \propto \ln \left\langle \mathcal{I}^{(N)}_L  \right\rangle$ ($s \gg 1$), yielding $\chi(L,W) = 0$ \cite{BookMehta}.  In the case of localized eigenfunctions, the energy levels are uncorrelated random variables with a Poisson distribution, the number variance is given by $n^{(N)}(L) =  \langle \mathcal{I}^{(N)}_L \rangle$ ($s \gg 1$) and, consequently, we have $\chi(L,W) = 1$ \cite{BookMehta}. Finally, if the energy levels follow a sub-Poisson distribution, the number variance scales linearly with $\langle \mathcal{I}^{(N)}_L \rangle$ ($s \gg 1$), but $0 < \chi(L,W) < 1$ \cite{Altshuler88,Chalker1996,Bogomolny2011,Mirlin}. This is the typical behavior of $\chi(L,W)$ at the critical point for the Anderson transition in finite dimensions \cite{Chalker1996} as well as
for critical random matrix ensembles \cite{Bogomolny2011}, in which the eigenfunctions exhibit a multifractal behavior. Thus, the level-compressibility is a suitable quantity to distinguish among Wigner-Dyson, Poisson and sub-Poisson statistics of the energy levels.

The first $\kappa_1^{(N)}$ and second $\kappa_2^{(N)}$ cumulants of the random variable $\mathcal{I}^{(N)}_L$ read
\begin{eqnarray}
\kappa_1^{(N)}(L,W)= \frac{\partial \mathcal{F}_L^{(N)}(y)}{\partial y}\Big|_{y=0} =  \frac{ \left\langle \mathcal{I}^{(N)}_L  \right\rangle}{N}  \,, \label{fgh0} \\
\kappa_2^{(N)}(L,W)= - \frac{\partial^2 \mathcal{F}_L^{(N)}(y)}{\partial y^2}\Big|_{y=0} =  \frac{n^{(N)}(L)}{N}   \,,
\label{fgh}
\end{eqnarray}
where the cumulant generating function $\mathcal{F}_L^{(N)}(y)$ for the statistics of $\mathcal{I}^{(N)}_L$ is given by
 \begin{equation}
 \begin{split}
\mathcal{F}_L^{(N)}(y)=-\frac{1}{N}\ln \bracket{e^{-y\mathcal{I}_L^{(N)}}} \,.
\label{eq:cgf1}
\end{split}
\end{equation}
Substituting eqs. (\ref{fgh0}) and (\ref{fgh}) in eq. (\ref{hhj}), we see that the level-compressibility can be written in terms of the cumulants 
\begin{equation}
\chi(L,W) = \frac{\kappa_2(L,W) }{\kappa_1(L,W)}\,,
\end{equation}
with $\kappa_{1,2}(L,W) \equiv \lim_{N \rightarrow \infty} \kappa_{1,2}^{(N)}(L,W)$. Thus, the calculation of $\chi(L,W)$ reduces to evaluate $\mathcal{F}_L^{(N)}(y)$ in the limit $N \rightarrow \infty$, from which the first and second cumulants are readily obtained.

Here we briefly recall the analytical approach to calculate $\lim_{N \rightarrow \infty} \mathcal{F}_L^{(N)}(y)$ and then we present the final equations  for the first and second cumulants. A detailed account of this computation is presented in \cite{Metz2016}. Our first task consists in expressing $\mathcal{F}_L^{(N)}(y)$ in terms of the disordered  Hamiltonian $\mathcal{H}$, such that we are able to compute analytically the ensemble average $\langle \dots \rangle$ in eq. (\ref{eq:cgf1}).  This is achieved by representing the Heaviside step function $\Theta(x)$ ($x \in \mathbb{R}$) in terms of the discontinuity of the complex logarithm along the negative real axis, i.e.,  $\Theta(-x)=\frac{1}{2\pi i}\lim_{\eta\to 0^{+}}[\ln (x+i\eta)-\ln (x-i\eta)]$. Using this prescription in eq. (\ref{ssq1}), we derive

 \begin{equation}
 \begin{split}
\mathcal{I}_L^{(N)}&=-\frac{1}{\pi i}\lim_{\eta\to0^{+}}\ln\left[\frac{Z( L/2-i\eta)Z(-L/2+i\eta)}{Z( L/2+i\eta)Z( -L/2-i\eta)}\right]
\label{eq:pin}\,,
\end{split}
\end{equation}
where  $Z(z)=\left[\det\left(\mathcal{H}-z\openone\right)\right]^{-1/2}$ ($z \in \mathbb{C}$), with   $\openone$ the $N\times N$ identity  matrix, $(\cdots)^\star$ the complex conjugation, and $\eta > 0$ a regularizer. Combining eqs. \eqref{eq:pin} and \eqref{eq:cgf1}, one can write 
\begin{equation}
 \begin{split}
\mathcal{F}_L^{(N)}(y)=- \lim_{\eta\to 0^+}\frac{1}{N}\ln \mathcal{Q}_L^{(N)}(y)\,,
\label{sda}
\end{split}
\end{equation}
with
\begin{equation}
 \begin{split}
\mathcal{Q}_L^{(N)}(y)&=\Big\langle Z^{\frac{i y}{\pi }}(L/2+i\eta) Z^{\frac{i y}{\pi }}(-L/2- i{\eta})\\
&\times Z^{-\frac{i y}{\pi }}(L/2 -i\eta)Z^{-\frac{i y}{\pi }}(-L/2+i{\eta}) \Big\rangle \,.
\label{eq:fq}
\end{split}
\end{equation}
The ensemble average in eq. (\ref{eq:fq}) can be calculated analytically using the replica approach of  spin-glass theory \cite{BookParisi,Metz2016}. The limit $N \rightarrow \infty$ of $\mathcal{F}_L^{(N)}(y)$ follows from the solution of a saddle-point integral, in which we have restricted our analysis to those saddle-points that are replica symmetric, i.e., invariant with respect to the permutation of two or more replicas. For all details involved in the calculation of
$\lim_{N \rightarrow \infty} \mathcal{F}_L^{(N)}(y)$, we refer the reader to the supplemental information of \cite{Metz2016}.

Following this approach we find the expressions for the first two cumulants: 
\begin{align}
\kappa_{1}(L,W) &= \frac{1}{\pi} \lim_{\eta \rightarrow 0^{+}} \left[  \frac{(k+1)}{2} \left\langle F   \right\rangle_{\nu} -  \left\langle R   \right\rangle_{\mu} 
- (k+1)  \left\langle R   \right\rangle_{\nu} \right]\,, \label{k1} \\
\kappa_{2}(L,W)&= \frac{1}{\pi^2} \lim_{\eta \rightarrow 0^{+}} \Bigg[   \left\langle R^2   \right\rangle_{\mu} - \left\langle R  \right\rangle^{2}_{\mu} \nonumber \\   
&+  2 (k+1) \left( \left\langle R F   \right\rangle_{\nu} - \left\langle R   \right\rangle_{\nu} \left\langle F  \right\rangle_{\nu}     \right)\nonumber\\ 
&- \frac{(k+1)}{ 2} \left( \left\langle F^{2}   \right\rangle_{\nu} - \left\langle F   \right\rangle^{2}_{\nu}    \right)      \nonumber \\
&- (k+1)  \left( \left\langle R^2   \right\rangle_{\nu} - \left\langle R   \right\rangle^{2}_{\nu}    \right) \Bigg]\,, \label{k2}
\end{align}
with the functions $R = R(u,v)$ and $F = F(u,v;u^{\prime},v^{\prime})$ 

\begin{align}
R(u,v) &= \frac{i}{2} \ln{\left[ \frac{u v }{\left(u v \right)^{*} } \right]}\,, \\
 F(u,v;u^{\prime},v^{\prime}) &= R(u,v) + R(u^{\prime},v^{\prime})\nonumber\\
 & + \varphi(u,u^{\prime}) + \varphi(v,v^{\prime})\,,
\end{align}
and 
\begin{equation}
\varphi(u,u^{\prime}) = - \frac{i}{2} \ln{\left[  \frac{ 1 +u u^{\prime}     }{ \left( 1 + u u^{\prime} \right)^{*}  }   \right]}\,.
\end{equation}
The average  $\left\langle \dots  \right\rangle_{\mathcal{P}}$ of integer powers of $R(u,v)$ and $F(u,v;u^{\prime},v^{\prime})$ with an arbitrary distribution $\mathcal{P}$ is  defined by the general formula
\begin{equation}
\begin{split}
\left\langle R^{m} F^{n}   \right\rangle_{  \mathcal{P} } &=  \int d u \, d v \, d v^{\prime} \, d u^{\prime} \, \mathcal{P}(u,v) \mathcal{P}(u^{\prime},v^{\prime})\\
&\times R^{m}(u,v) F^{n}(u,v;u^{\prime},v^{\prime}) \,,
\label{gqp}
\end{split}
\end{equation} 
where $m \geq 0$ and $n \geq 0$. The distributions $\mu(u,v)$ and $\nu (u,v)$, which enter in the averages appearing in eqs. (\ref{k1}) and (\ref{k2}), are determined from the self-consistent equations
\begin{widetext}
\begin{align}
\mu(u,v) &= \int \left( \prod_{r=1}^{k+1} du_r \,  dv_r \, \nu(u_r,v_r)  \right)
\left\langle  \delta\left[ u - \frac{1}{i \left( \epsilon - \frac{L}{2}-i\eta \right) - \sum_{r=1}^{k+1} {u_r}} \right]    
\delta\left[ v + \frac{1}{i \left( \epsilon + \frac{L}{2}+i\eta \right) - \sum_{r=1}^{k+1}{v_r}} \right]  
 \right\rangle_{\epsilon} ,
\label{1tra} \\
\nu(u,v) &= \int \left( \prod_{r=1}^{k} du_r \,  dv_r \, \nu(u_r,v_r)  \right)
\left\langle  \delta\left[ u - \frac{1}{i \left( \epsilon - \frac{L}{2}-i\eta \right) - \sum_{r=1}^{k} {u_r}} \right]    
\delta\left[ v + \frac{1}{i \left( \epsilon + \frac{L}{2}+i\eta \right) - \sum_{r=1}^{k} {v_r}} \right]  
 \right\rangle_{\epsilon} .
\label{2tra}
\end{align}
\end{widetext}
where $\langle \dots \rangle_{\epsilon}$ is the average over the on-site random potentials.

Equations (\ref{k1}-\ref{2tra}) are exact for $N\to\infty$ and $L =\mathcal{O}(1)$ fixed, independently of the system  size $N$, and the level-compressibility is evaluated with high accuracy using  the population dynamics algorithm \cite{Metz2016}. However, as we discussed above, we should calculate $\chi(L,W)$ at the scale $L=\mathcal{O}(1/N)$ in order to probe the statistics of the energy levels within the extended phase. The reason is twofold: (i) by considering the regime  $L=\mathcal{O}(1/N)$, we are inspecting the fluctuations of low-lying energies of $O(1/N)$ (or  long time scales of the order $O(N)$, much larger than the typical size $\ln N$ of the loops); (ii) the average density of states $\langle \rho(E) \rangle$ is uniform along an interval of size $L=\mathcal{O}(1/N)$, and we avoid the spurious influence on $\chi(L,W)$ of significant variations of $\langle \rho(E) \rangle$ \cite{Metz2016}.

In principle, one should employ the formalism of \cite{Metz2016} and determine the cumulants when $L=s/N$ ($s \gg 1$). However,  one immediately concludes that the terms arising  from rescaling $L \rightarrow s/N$ and $\eta \rightarrow \eta/N$ in the formal development \cite{Metz2016} enter only in an eventual calculation involving finite size corrections, i.e., when one considers $N$ large but finite  \cite{MetzPar,Metz2015b}. Thus, the leading behavior of the level-compressibility $\lim_{N \rightarrow \infty} \kappa_2^{(N)}/\kappa_1^{(N)}$ in the scaling regime $L=O(1/N)$ should already be given by eqs. (\ref{k1}-\ref{k2}) in the limit $L \rightarrow 0^{+}$ and $\eta \rightarrow 0^+$. The central idea here is to explore numerically this limit using population dynamics. Note that the imaginary part of  the energy $\eta$ is also going to zero and  the interesting limit is $L \to 0^{+}$ and $\eta \to 0^+$, keeping the ratio $L/\eta$ large. Essentially, $\eta$ plays the  role  of the level spacing in a regularized density of states and we want to have many levels within $[-L/2,L/2]$.

Due to the $\eta$-dependency of eqs. (\ref{k1}-\ref{2tra}),  it is convenient to introduce the level compressibility $\chi_\eta(W,L)$ for fixed  $\eta$. For a given value of $L$ and $W$, we have
\beeq{
\chi(W,L)=\lim_{\eta\to 0^{+}}\chi_\eta(W,L)\,.
\label{ss1}
}
Henceforth, we restrict ourselves to $k=2$. For this connectivity, the eigenfunctions at the center of the band undergo an Anderson localization transition at $W_c = 17.4$  \cite{Abou73,Biroli2010}.

\begin{figure}
\begin{center}
\includegraphics[width=8.5cm,height=6cm]{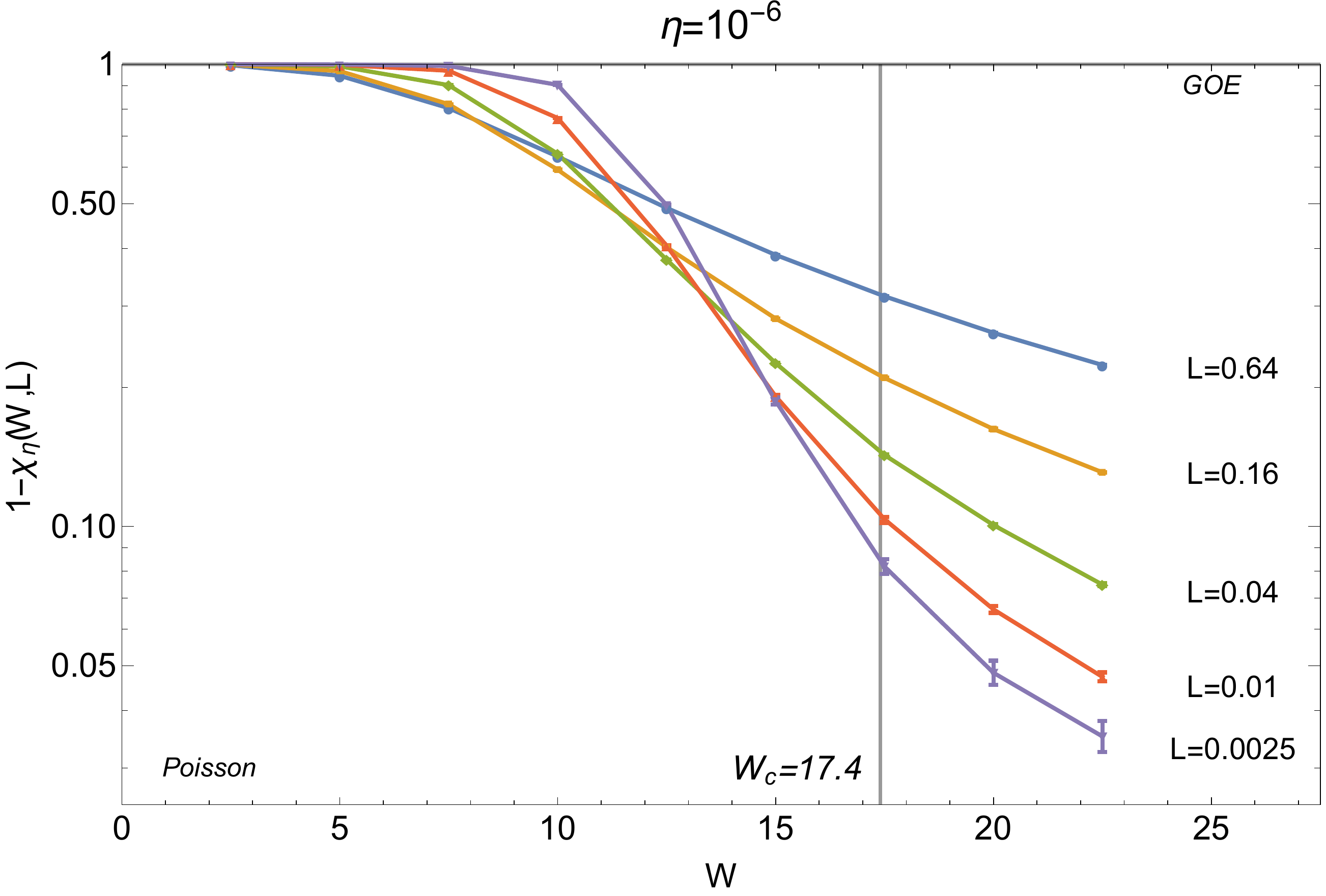}
\caption{The level-compressibility as a function of $W$ for a fixed imaginary energy $\eta=10^{-6}$
and different sizes of the interval $[-L/2,L/2]$. The number of neighbors connected to each node is $k+1 = 3$.}
\label{fig1}
\end{center}
\end{figure}

In figure \ref{fig1} we present results for $\chi_\eta(W,L)$ as a function of $W$ for fixed $\eta=10^{-6}$ and different values of the size $L$ of the interval. As $L$ decreases, it is clear that $\chi_\eta(W,L)$ converges to $\chi_\eta(W,L)=1$ or $\chi_\eta(W,L)=0$ for $W > W_c$ or $W < W_c$, respectively, as long as $W$ is not too close to the critical value $W_c = 17.4$.
Importantly, one observes a non-monotonic behavior of $\chi_{\eta}(W,L)$ as a function of $L$ for  some values of $W < W_c$. This is particularly evident for $W=10$ and $W=12.5$. However, it is not clear from figure \ref{fig1} that  $\eta$ is small enough such that the limit $\eta \rightarrow 0^+$ has been reached, especially close to the critical point.

In order to have reliable data in the delocalized phase $W< W_c$, it is crucial to understand the dependence of $\chi_\eta(W,L)$ with  respect to $\eta$. We have thus solved eqs. (\ref{k1}-\ref{2tra}) for several values of $\eta$, keeping $L$ fixed. For sufficiently small $\eta<\eta_{*} \sim L$, $1 - \chi_\eta(W,L)$ can be well fitted by the function $\chi_0+a \eta^{b}$, where the fitting parameters $\chi_0(W,L)$, $a(W,L)$ and $b(W,L)$ can be determined with high accuracy (see the appendix). This procedure allows to obtain $\chi(W,L)=\lim_{\eta\to 0^+}\chi_{\eta}(W,L)$  simply by reading the value of $\chi_0$.  Performing this numerical computation for many values of $W$ and $L$ is highly  time consuming, so we have focused on some values of $W$ within the  extended phase for which the eigenfunctions would be multifractal, according to recent works \cite{Luca2014,Alt2016, Altshuler2016}.

\begin{figure}[h]
\begin{center}
\includegraphics[width=8.5cm,height=6cm]{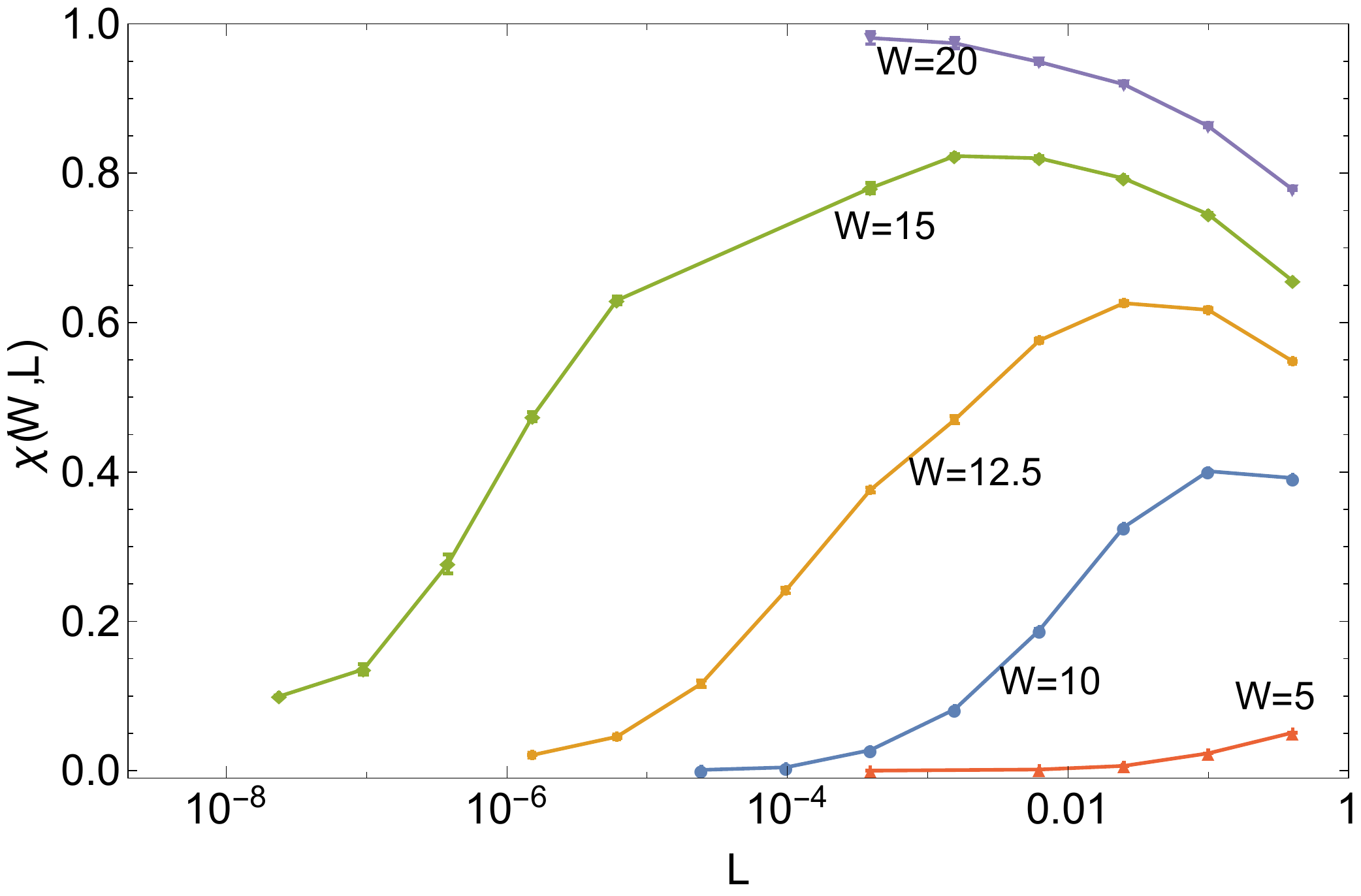}
\caption{The behavior of the level-compressibility $\chi(W,L)$ as a function of $L$ for connectivity $k+1=3$ and different values of the disorder strength $W$.  For $W< W_c \simeq 17.4$, the function $\chi(W,L)$ approaches zero in the limit $L \rightarrow 0^+$, signaling the Wigner-Dyson statistics of the energy levels.}
\label{fig3}
\end{center}
\end{figure}

The main outcome of this calculation is shown in figure \ref{fig3}, where we show $\chi(W,L)$ as a function of $L$.
As we approach $W_c$ from the delocalized side, the level compressibility  $\chi(W,L)$ behaves non-monotonically as a function of $L$: initially it tends to its Poisson value $\chi(W,L) =1$,  but as $L$ is further decreased, the   level compressibility clearly moves towards the limit  $\lim_{L \rightarrow 0^{+}}\chi(W,L) = 0$, characteristic of Wigner-Dyson statistics \cite{BookMehta}.  As the critical point is approached, the regime where  $\chi(W,L)$ attains its maximum value sets in for smaller and smaller $L$, making the numerical calculation highly demanding.  In spite of this numerical difficulty, our results strongly indicate that $\lim_{L\to 0^{+}} \chi(W,L)= 0$ for $W < W_c$, supporting the Wigner-Dyson statistics of the energy levels in the whole extended phase.

\section{Discussion} \label{discussion}
 
In this work we have calculated the level-compressibility $\chi(W,L)$ of the energy levels within a box of size $L=O(1)$ for the Anderson model on an infinitely large regular random graph (RRG). We have focused on the behavior of $\chi(W,L)$ for $L \rightarrow 0^{+}$, from which we expect to characterize the statistics of the energy levels (Poisson, sub-Poisson or Wigner-Dyson) at a local scale, i.e., when $L=O(1/N)$. This expectation is confirmed by the behavior of $\lim_{L \rightarrow 0^{+}} \chi(W,L)$ {\it away from the critical point} $W_c$: we have found that $\lim_{L \rightarrow 0^{+}} \chi(W,L)$ converges to one or zero, provided $W > W_c$ or $W < W_c$, respectively. Hence we have employed the level-compressibility to probe the eigenvalue statistics as we approach the critical point $W_c$ from the delocalized phase. Our results show that, {\it for values of $W$ closer to the critical point $W_c$}, $\chi(W,L)$  approaches zero in the limit $L \rightarrow 0^{+}$. This is consistent with earlier analytical predictions for the Anderson model on Erd\"os-R\'enyi random graphs \cite{Mirlin91} as well as with recent numerical results for the Anderson model on regular random graphs \cite{Mirlin2016,Tikhonov2016}. Our results are free of finite size effects, since they are obtained from a set of exact equations valid in the limit $N \rightarrow \infty$.

In particular, we do not observe any change of behavior of $\lim_{L \rightarrow 0^+}  \chi(W,L)$ for $W < W_c$ (see figure 2), as one would expect from the results for the fractal exponents \cite{Alt2016,Altshuler2016}, in combination with the standard view according to which the level-compressibility at the scale $L = O(1/N)$ is directly related with the statistics of the eigenfunctions \cite{Chalker1996,Bogomolny2011}. From this perspective, our results support the ergodicity of the eigenfunctions within the whole extended phase, entirely consistent with numerical diagonalization results \cite{Mirlin2016}. On the other hand, recent works \cite{Kravtsov,Biroli1,Amini} show that, in the Rosenzweig-Porter model, the non-ergodic extended states are unveiled by considering the statistics of the eigenfunctions at the scale of the Thouless energy $E_{th} \sim N^{1-\gamma}$ ($1 < \gamma < 2$), much larger than the mean level spacing $1/N$. Since the Anderson model on a RRG belongs, in a certain sense, to the same class as the Rosenzweig-Porter model \cite{Kravtsov}, the sole results for $\chi$ seem not sufficient to conclude about the ergodicity of the eigenfunctions within the extended phase. As a future perspective, it would be interesting to extend our method in order to compute $\chi$ at the scale $L =O(E_{th})$. 

Recently it has been put forward that replica symmetry breaking should be taken into account to correctly describe the eigenfunctions in the extended phase \cite{Altshuler2016}. Equations (\ref{k1}-\ref{2tra}) are derived assuming replica symmetry, which is exact provided  we fix $L=O(1)$ for $N \rightarrow \infty$ \cite{Metz2016,Metz2015}. This is corroborated by an abundance of works \cite{Perez2008,Kuhn2008,Ergun,Metz2010,Metz2015}, where observables related to the global density of states $\langle \rho(E) \rangle$ of the adjacency  matrix of several treelike random graphs are calculated using replica symmetry and confirmed through direct diagonalization.

 Nevertheless, the issue of replica symmetry breaking could arise in the limit $L \rightarrow 0^{+}$, since we expect to approach the local scale of $L=s/N \, (s \gg 1)$ characterizing the mean level spacing. From the replica calculation of the connected part of the two-level correlation function $R^{(c)}(s)$ for the GUE ensemble \cite{Kamenev}, the replica-symmetric saddle-point yields the decay $R^{(c)}(s) \propto 1/s^2$, which already gives the leading contribution $\chi(W,s/N) \propto s^{-1} \ln s \overset{s \rightarrow \infty}{\longrightarrow} 0$ \cite{BookMehta,Mirlin}. The inclusion of saddle-points that break replica symmetry allow to capture the oscillatory behavior of $R^{(c)}(s)$ \cite{Kamenev}, which does not affect the leading value $\lim_{s \rightarrow \infty} \chi(W,s/N)=0$, but only introduces sub-leading corrections due to finite $s$. We expect the situation to be similar for the Anderson model on regular random graphs, i.e., replica symmetry breaking is important only when one is interested in sub-leading corrections for finite $s$. This is also supported by the absence of many solutions to the cavity or population dynamics eqs. (\ref{1tra}) and (\ref{2tra}), which is a common sign of replica symmetry breaking \cite{BookParisi}.

\vfill

\begin{acknowledgments}
 The authors thank Alexander Mirlin and Yan V. Fyodorov for illuminating discussions. FLM thanks the hospitality of the Institute of Physics at UNAM. The authors thank the support of DGTIC for the use of the HP Cluster Platform 3000SL, codename Miztli, partly done under the Mega-project  LANCAD-UNAM-DGTIC-333. We also thank Cecilia Nogu\'ez for letting us use the  computer cluster Baktum during the initial stages of this project. This work has been funded by the program UNAM-DGAPA-PAPIIT IA101815. 
\end{acknowledgments}

\appendix*

\section{The numerical procedure to calculate the level-compressibility}

In this appendix we explain how to evaluate the limit $L \rightarrow 0^+$ of the level-compressibility $\chi(W,L)$.  The function $\chi(W,L)$ for a given $L$ is obtained by considering the limit $\eta \rightarrow 0^{+}$ of eqs. (\ref{1tra}) and (\ref{2tra}), so that it is convenient to define the  level compressibility $\chi_\eta(W,L)$ for fixed $\eta$. 
The idea is to compute $\chi_\eta(W,L)$ for decreasing $\eta$. After we have obtained $\chi_\eta(W,L)$ for several values of $\eta$, we have done a 
non-linear fitting to extract $\chi(W,L)$ according to eq. (\ref{ss1}). 
\begin{figure}[H]
\centering
\includegraphics[scale=0.32]{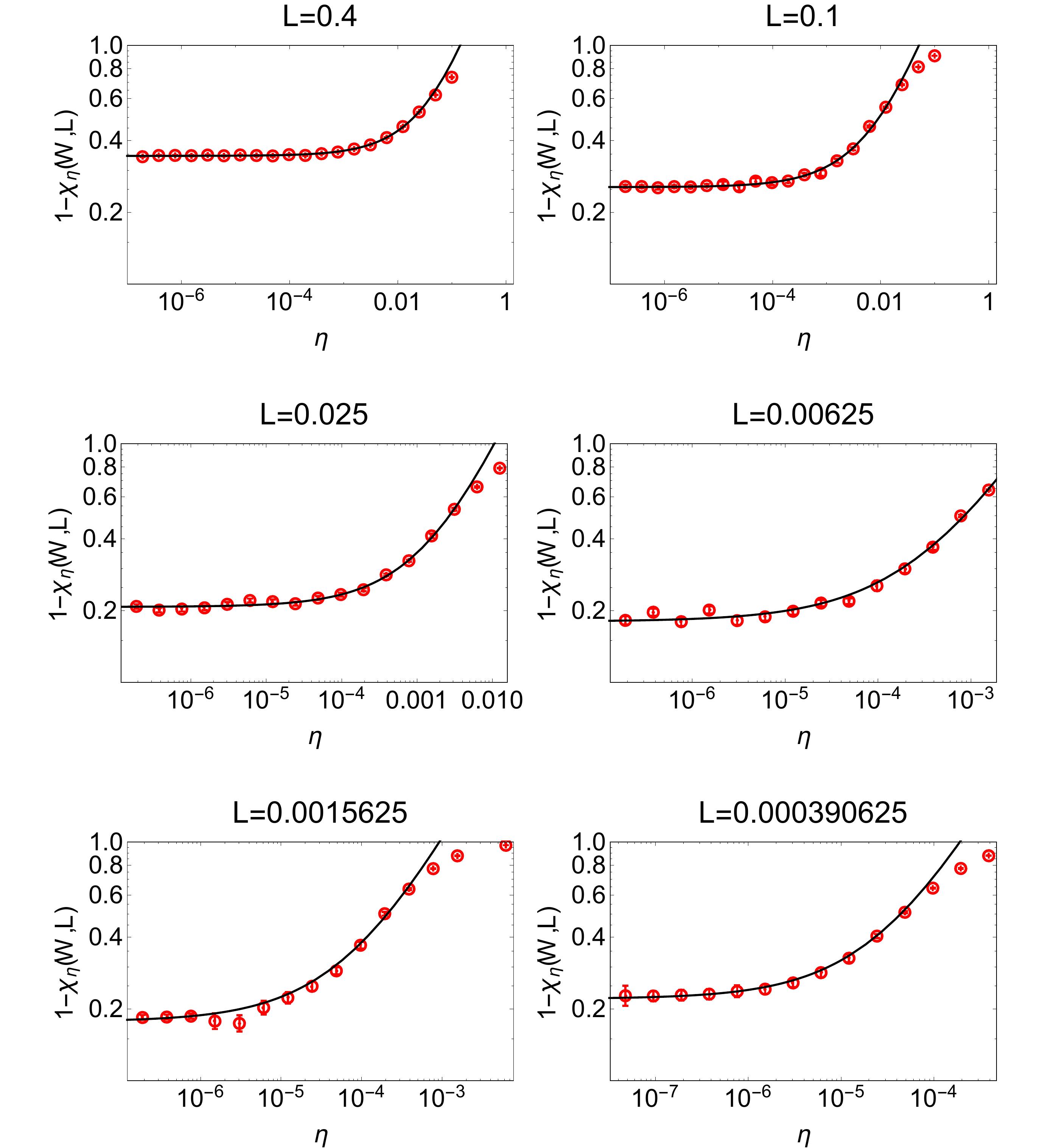}
\caption{Level compressibility $\chi_\eta(W,L)$ as a function of $\eta$ for $W=15$ and several values of $L$. The solid lines are the best fits of the numerical data
with the function $1-\chi_\eta(W,L)= \chi_0(W,L)+a (W,L) \eta^{b(W,L)}$. The values of the fitting parameters can be found in Table \ref{table1}.}
\label{ffig2w15}
\end{figure}

Equations (\ref{1tra}) and (\ref{2tra}) of the main text are solved numerically using the population dynamics algorithm, in which the distributions $\mu(u,v)$ and $\nu(u,v)$ are parametrized by a large number $\mathcal{N}$ of random variables, consistently updated according to eqs. (\ref{1tra}) and (\ref{2tra}). We point out that $\mathcal{N}$ serves only to represent the continuous distributions
appearing in eqs. (\ref{1tra}) and (\ref{2tra}) in terms of a population of random variables, and the parameter $\mathcal{N}$ does not have in principle any relation with
the system size $N$. In fact, eqs.  (\ref{1tra}) and (\ref{2tra}) are valid strictly in the limit $N \rightarrow \infty$, by taking the ensemble average 
over the regular random graph realizations and over the diagonal random potentials.
We refer the reader to \cite{Metz2016} for a detailed account of population dynamics.  

For a fixed value of $\eta$, $W$ and $L$, we consider a very large population size $\mathcal{N}$ between $10^{7}$  and $10^{8}$, we solve eqs. (18) and (19) using population dynamics,  and we obtain a value for $\chi_\eta(W,L)$. Since  $\chi_\eta(W,L)$ fluctuates among different realizations of population dynamics, we repeat this calculation several times, generating a dataset containing between $10^3$ to $10^4$ values of $\chi_\eta(W,L)$. Reliable estimates for the average value of $\chi_\eta(W,L)$ and for the corresponding error bar are derived from this dataset. For a fixed $W$ and $L$, we then evaluate $\chi_\eta(W,L)$ for several $\eta$ using this procedure, which allows us to determine very accurately the limit $\lim_{\eta \rightarrow 0^+} \chi_\eta(W,L)$.

We exemplify our numerical approach in figure \ref{ffig2w15}, where we show the outcome of our numerical calculations for $W=15$ and different values of $L$. The limit $\eta \rightarrow 0^+$ of $\chi_\eta(W,L)$ in each graph is obtained by noticing that, for some $\eta < \eta_{*} \sim L$, the function $1-\chi_\eta(W,L)$ depends on $\eta$ according to the power-law $1-\chi_\eta(W,L)= \chi_0(W,L)+a (W,L) \eta^{b(W,L)}$. Table \ref{table1} reports the fitting parameters obtained by performing a non-linear least-squares fitting with the function $1-\chi_\eta(W,L)= \chi_0(W,L)+a (W,L) \eta^{b(W,L)}$, using the program \texttt{gnuplot}. The asymptotic behavior $\chi(W,L) = \lim_{\eta \rightarrow 0^+} \chi_\eta(W,L)$ is simply obtained from the value of $\chi_0(W,L)$ as $\chi(W,L)=1-\chi_0(W,L)$.

\begin{table}[h]
\begin{center}
\begin{tabular}{ |c|c|c|c|c|c| } 
 \hline
 $L$ & $\chi_0(W,L)$ &$a(W,L)$&$b(W,L)$ & ${\rm ndf}$ & $\chi^2/{\rm ndf}$ \\ \hline
0.4&0.3442(6) &2.7(1) & 0.73(1) &15 & 1.32\\ 
0.1&0.255(2) &5.3(5) &0.56(2) & 14 & 1.03\\ 
0.025&0.207(2) &21(2) &0.73(2) & 12 & 1.42\\ 
0.00625&0.180(3) &27(3) &0.63(2) & 11 & 2.59\\ 
0.0015625&0.177(3)  &66(13) & 0.63(3) & 9 & 1.81\\ 
0.000390625& 0.220(7)  &312(184) & 0.70(6) & 8 & 0.15\\ \hline
\end{tabular}
\end{center}
\caption{Numerical estimates for the parameters obtained from the non-linear fitting of the data in figure \ref{ffig2w15} with the function 
 $1-\chi_\eta(W,L)= \chi_0(W,L)+a (W,L) \eta^{b(W,L)}$. The parameter ${\rm ndf}$ is the number of degrees of freedom, while
$\chi^2$ is the standard function to be minimized in the non-linear squares fitting ($\chi^2$ measures the deviation of the dataset from the 
analytical curve and it should not be confused with the level-compressibility).}
\label{table1}
\end{table}

\bibliography{biblio.bib}

\end{document}